# Review
# Komatiites: From Earth's Geological Settings to Planetary and Astrobiological Contexts

Delphine Nna-Mvondo[1] and Jesus Martinez-Frias[1]

[1] *Planetary Geology Laboratory, Centro de Astrobiologia (CSIC/INTA), associated to NASA Astrobiology Institute, Ctra. De Ajalvir, km 4. 28850 Torrejon de Ardoz, Madrid, Spain.*

**Correspondence:**

Laboratorio de Geología Planetaria, Centro de Astrobiología (CSIC/INTA), associated to NASA Astrobiology Institute, Instituto Nacional de Técnica Aeroespacial, Ctra. De Ajalvir, km 4. 28850 Torrejón de Ardoz, Madrid, Spain.

Phone: +34 915206434

Fax: +34 915201074

E-mail: nnamvondod@inta.es

## ABSTRACT

Komatiites are fascinating volcanic rocks. They are among the most ancient lavas of the Earth following the 3.8 Ga pillow basalts at Isua and they represent some of the oldest ultramafic magmatic rocks preserved in the Earth's crust at 3.5 Ga. This fact, linked to their particular features (high magnesium content, high melting temperatures, low dynamic viscosities, etc.), has attracted the community of geoscientists since their discovery in the early sixties, who have tried to determine their origin and understand their meaning in the context of terrestrial mantle evolution. In addition, it has been proposed that komatiites are not restricted to our planet, but they could be found in other extraterrestrial settings in our Solar System (particularly on Mars and Io). It is important to note that komatiites may be extremely significant in the study of the origins and evolution of Life on Earth. They not only preserve essential geochemical clues of the interaction between the pristine Earth rocks and atmosphere, but also may have been potential suitable sites for biological processes to develop. Thus, besides reviewing the main geodynamic, petrological and geochemical characteristics of komatiites, this paper also aims to widen their investigation beyond the classical geological prospect, calling attention to them as attractive rocks for research in planetology and astrobiology.

**Abbreviations:** BGB: Barberton Greenstone Belt; GI: Gorgona Island; HSE: High Siderophile Element; LREE: Light Rare Earth Element; PGE: Platinum Group Element; PH: Pyke Hill; REE: Rare Earth Element; VB: Vetreny Belt

# 1. Introduction

Astrobiology is a quickly evolving, interdisciplinary field of science. Among the research for understanding how life began and evolved on Earth, various studies have increased evidence that early life may have been connected to volcanic settings. Hence, it is important to study and constrain the volcanic environment and rocks.

In the early days of the Earth, evidence shows that the planet had much more energetic activity than today (as in the case on other terrestrial planets). A natural consequence of this state would have been a higher mantle temperature and higher degree of melting, leading to the production of lavas with compositions closer to the bulk mantle chemical structure than those today. In this idea, "primitive" lavas have been considered as those formed at the highest temperatures. Such odd lavas do exist in the terrestrial geological record and are identified as komatiites. Komatiites were first recognized in the late 1960s in the Barberton Mountainland greenstone belt in South Africa (Viljoen and Viljoen, 1969a, 1969b) and were named from their type locality along the Komati River. In the beginning, it was thought that komatiite eruptions appeared mainly in the Archean. Nonetheless, this assumption is incorrect as some have been dated to the Paleozoic and Mesozoic ages. Also it is hypothesized that, in the late Archean, komatiite volcanoes would have built edifices on the surface forming oceanic plateaux. These plateaux would be unsubductable and would become part of the land. Earlier in the Archean (4.0 – 4.2 Ga), mid-ocean ridges themselves may have been komatiitic. As time went on, the ridges changed to erupting basalt, as they do today, but hotspots continued to produce komatiites until the end of the Archean (Shock, 1996).

Although the typology and classification of komatiites has been terminologically debated (refer to detailed information described by Malyuk and Sivoronov, 1984), komatiites are petrologically defined as ultramafic extrusive volcanic rocks, which derived from highly magnesian magmas. Their MgO contents exceed 18 wt% compared to 10 - 15 wt% for the majority of mafic rocks. From Viljoen and Viljoen (1969b), they are classified in two groups, based on their composition, the basaltic komatiites (18 – 24 % MgO) and the peridotitic komatiites (> 24 % MgO). Also, they are noted for their low dynamic viscosities (0.1 to 1 Pa·s), high liquidus temperature (~1600 °C) and great potential for turbulent flow and thermal erosion of their underlying substrates. Although their origin is still controversial, it is generally accepted that komatiites were generated most probably at depth of 150 to 200 km by massive partial melting of the Archean mantle (Takahashi and Scarfe, 1985) and that the ancient komatiitic lava flows erupted at high temperatures of 1400 - 1700 °C (Arndt *et al.*, 1979; Huppert *et al.*, 1984). In comparison, anhydrous liquidus temperatures (at 1 bar pressure) of Hawaiian tholeiitic picrites average 1365°C (Green *et al.*, 2001). Komatiites are exceptional and it can be said that, as a whole, they have no cenozoic analogs.

The origin of komatiitic Archean volcanism has also been discussed ( see details in chapter 3), considering, among the various hypothesis, a possible association with meteorites and meteoritic impacts (Green, 1972; Jones, 2002; Jones *et al.*, 2003). This argument supports the plausible scenario of an increased supply of extraterrestrial material to the primitive Earth and perhaps also to other planets and satellites of our solar system (Fodor and Keil, 1976; Bairds and Clark, 1981; Williams *et al.*, 2000a, b; Kargel *et al.*, 2003; Rosengren *et al.*, 2004).

Until now, komatiites have been mainly studied for their singular geochemical and tectonic characteristics and significance. The volume of data collected in this field of investigation, particularly regarding their petrology, is quite impressive. An updated search on komatiites using the web of Science (of the ISI web of knowledge website), gives a result of more than 500 published papers in thirty years. And during these thirty years, the last 5 years have seen the largest number of publications per year showing a vigorous recent interest in komatiites. Recently, the study of komatiites has also been extended to planetary geology, as they could be of a great help in understanding igneous processes outside the Earth on other planets and moons, where analog materials could exist (Venus, Mars, Io and lunar lavas). In this paper, we are synthesizing and discussing the information reported about komatiites, also stressing their possible occurrence in other extraterrestrial environments and their astrobiological significance.

# 2. Age

As previously defined, komatiites have mostly an early origin. Textural and chemical analyses of more than 20,000 rocks have revealed that peridotitic komatiite is at present unknown from modern environments and basaltic komatiite is rare (Brooks and Hart, 1974). Komatiites were produced most commonly during the Archean (>2.7 Ga) and late Archean. Few are found in the early Proterozoic and they were much less frequent in the Phanerozoic era. The prevalence of komatiite almost entirely in the Archean indicates fundamental differences between ancient and modern mantle conditions. One of the reasons for the decline of the abundance from the Archean to the Phanerozoic is interpreted as a result of the decrease in the degree of melting due to secular cooling of the mantle since the Archean (Nisbet *et al.*, 1993). As a consequence, this interpretation makes komatiites as potential thermometers of the Earth's cooling. It has also been proposed that the deep mantle sources that produced komatiites have undergone chemical changes over the geological time giving rise to more enriched magmas (Campbell, 1998).

In Table 1, we report the age of komatiites at representative selected locations. Komatiites are encountered in a number of places around the world: Canada, Southern African, Australia, Baltic Shield, Colombia, and Vietnam, among others. Geochronologic dating shows the Barberton komatiites in South Africa to be 3.6 - 3.2 billion years old, which represents some of the oldest ultramafic rocks found on Earth (extrusive rocks with an age of ~3.8 Ga crop out at Isua in Greenland). Such "timing" fits nicely with the concept of a hot early Earth taking into account the high temperatures inferred for the komatiite source region. Then there is a prevalence of komatiites around 2.7 Ga, most of them located in Canada and Australia. Munro Township, Ontario, Canada is the best preserved komatiite location in the world. Although komatiites occur in Proterozoic formations (e.g. Finnish Lapland), they are not as common as in Archean greenstone belts. Komatiites are very rare in the Phanerozoic geological record. The most prominent representatives are the Mesozoic komatiites from Gorgona Island in Colombia associated with the Caribbean oceanic plateau (Echeverria, 1980; Arndt *et al*., 1997). They are the youngest known komatiitic lavas (89 Ma) and their existence has been interpreted by some authors (Brandom *et al.,* 2003) as unusually high temperatures in the mantle. Nevertheless, caution is needed because for example seismic data indicate that the average oceanic crust in the mid-Cretaceous could not be thicker than it is now. The occurrence of younger komatiites is less common than the Barberton and Munro-types. One of the explanations has been related to a progressive dehydration of the upper mantle in the mid- to late-Archean. In this assumption, the Mesozoic appearance of komatiites may be due to local hydration of the upper mantle after dehydration of hydrous phases (Inoue *et al*., 2000).

[TABLE 1]

# 3. Origin

The origin of the anomalously high-temperature komatiitic magmas has been debated for many years because of the implications for the thermal structure and composition of the Archean Earth mantle. It is a very controversial issue and many hypotheses have been proposed and discussed since their first recognition in the late 1960s, trying to find a model which best fits komatiites.

Komatiites were first explained as a result of a catastrophic melting event triggered by convective overturn during core formation (Viljoens and Viljoens, 1969a). However, this idea has been rapidly excluded, at least in the sense with which it was put forward.

Later on, a second scenario was proposed arguing a meteoritic impact origin. Spinifex (see next chapter) komatiites were long ago suggested to be an impact melt from chondritic meteorites; in this sense, it is important to note that cosmic spherules also feature spinifex olivine (Green, 1972; Jones, 2002). In such cases, ultramafic liquids from some Archean greenstone belts are interpreted as products of 60 - 80% melting of their mantle source composition. This would imply more catastrophic conditions of mantle melting than in Phanerozoic mantle dynamics, as a possible consequence of major impacts on the early Earth's surface. Therefore, following this hypothesis, Archean greenstone belts could correspond to very large impact scars, initially filled with impact-triggered melts of ultramafic to mafic composition and thereafter evolving with further magmatism, deformation and metamorphism (Green, 1972). Although this meteorite impact-related origin of komatiites still remains controversial, it has been recently re-assessed and more strongly supported (Abbott, 2000; Abbott and Isley, 2002). A temporal relationship between large impacts on the early Earth and Moon and mantle plume activities has been suggested by spectral analysis of time series of mantle plume rocks and impacts over the last 3.8 Ga (Abbott, 2000). Similarities between spectra derived from the impacts and the plume material could be due to strengthening of existing plume by the seismic energy released during impacts (Abbott and Isley, 2002). Such speculation could help to explain, for instance, why komatiites are not always plume tails and why they were more abundant during the Archean when large meteorites were frequently bombarding the early Earth. In this context, large meteorite and cometary impacts would have increased the amount of volcanism from already active mantle plumes. Recently, it has been pointed out that large impact events were not limited to the Hadean. There is an emerging body of evidence that the Archean geologic record, 3.5-2.5 Ga, preserves a signature of continuing bombardment (Lowe *et al.,* 2003).

The most widely accepted hypothesis about komatiite origin states that they were formed in an Archean plume-dominated environment (Fyfe, 1978; Arndt and Nisbet, 1982). Campbell *et al.* (1989) have argued from fluid-dynamics calculations that both basalts and komatiites could have been produced by a starting thermal plume rising in a warmer Archean mantle. In such a scenario, komatiites could form by melting in the hot axial jet of the starting plume, whereas basalts would be produced by melting in the large head of the plume in which cooler mantle is entrained. Komatiitic magmas generated by mantle plume activities could contribute to the formation of Archean oceanic plateaux which, in some cases, could be later on buried in the mantle via subduction (Polat *et al.*, 1998; Kerrich *et al.*, 1999; Puchtel *et al.*, 1999; Polat and Kerrich, 2000). They could also erupt on the continental crust as in the case of Zimbabwe komatiites (McDonough and Ireland, 1993), or interact with an island arc in a subduction environment (Wyman and Kerrich, 2002).

Another strongly supported scenario for a komatiite source is generation in a subduction zone. Brooks and Hart (1972, 1974) first pointed out that the major element chemistry of many of the komatiites and related magmas (komatiitic basalts) were more similar to modern mafic subduction magmas than to any magma thought to be produced by a modern plume. The first strong evidence for a subduction origin for komatiites was exposed from a study of the Nondweni komatiites (Wilson and Versfeld, 1994). These komatiites have much higher $SiO_2$ than the Munro or Gorgona rocks and show some similarities to modern mafic subduction magmas (boninites). High $SiO_2$ does not match with the plume scenario as $SiO_2$ contents of magmas generally decrease as the pressure of melting increases. Boninites have high $SiO_2$ at high MgO contents because they are high degree melts that are formed at shallow depths. Such melting can occur at shallow depth because it is caused by high contents of water (Crawford *et al.*, 1989). Boninites form by hydrous melting of metasomatized mantle above a subduction zone. A subduction zone origin has also been proposed for the Barberton komatiites and basaltic komatiites, because it appears to be more consistent with the available trace element data. Moreover plume-based models that appeal to majorite garnet fractionation cannot produce the low Ti/Zr ratios of komatiites or their wide variation in La/Sm (Parman *et al.*, 2003).

Because of their similarities to boninites, the old model (Allègre, 1982) of the production of komatiites by hydrous melting processes in the upper mantle has been reconsidered (Parman *et al.*, 2001). Hanski (1992) discovered pargasitic amphibole in Fe-rich komatiites in Finland. Stone *et al.* (1997) also found this hydrous amphibole in a komatiite. This mineral was probably the first and most significant evidence for high $H_2O$ contents. Nevertheless, there are more proofs suggesting that the Barberton-type komatiites were formed in the presence of water (see for instance Stone *et al.*, 1995, 1997; Parman *et al.*, 1997). The subduction zone origin focuses on the high $SiO_2$ content in komatiites whereas the plume model implies low $SiO_2$. A very detailed geochemical analysis (major and trace elements) comparing basaltic komatiites and modern mafic magmas was performed by Parman *et al.* (2001, 2003). Their results confirm that, excluding the Mg content, the compositions of the basaltic komatiites are very close to modern boninites and much more similar than any ocean island basalts.

With such geochemical resemblances, komatiites are strongly supposed to be produced by similar melting processes as for modern boninites but under hotter mantle temperature since one assumed that the Archean mantle was 100 - 500 °C hotter than the modern mantle (Parman *et al.*, 2001). Experimental data indicates that the Archean sub-arc mantle needs only to be 1500 - 1600 °C to produce hydrous komatiitic melts (Parman *et al.*, 2001). This is considerably cooler than estimates of mantle temperatures assuming an anhydrous, plume origin for komatiites (up to 1900 °C).

It is still unclear whether the water found in komatiites is produced in a subduction zone or whether it originated from a hydrous plume. For this reason, the water content of komatiites still remains a debated issue. Arndt *et al.* (1998) have compiled all the information regarding the hydrous/anhydrous origin debate for komatiites in an excellent review. In short, arguments supporting the hypothesis of the formation of komatiites by hydrous melting are the following: a) the presence of water in the mantle source of komatiites reduces the melting temperatures from very high to lower values (Allègre, 1982; Inoue, 1994; Kawamoto *et al.*, 1996); b) the pyroclatiscity and vesicularity containing features of some komatiites are characteristic of magmatic volatiles (de Wit *et al.*, 1983, 1987; Parman *et al.*, 1997), and c) experimental studies of peridotite melting argue that the chemical composition and the spinifex textures of komatiites require hydrous conditions (Inoue, 1994; Grove *et al.*, 1996; Ohtani *et al.*, 1997; Parman *et al.*, 1997). On the other hand, the scenario of melting within hydrous mantle is counterargued by: a) when hydrous komatiites are close to the surface, the loss of volatiles should produce degassing structures and textures which are rare in komatiites, and b) chemical and isotopic compositions of most komatiites indicate that their mantle source became depleted in incompatible elements soon before the formation of the magma, which may remove water. The existing experimental data are not yet conclusive, therefore until further work one may assume that most komatiites form in unusually hot and dry parts of the mantle, and only some rare komatiites are hydrous (Arndt *et al.*, 1998).

# 4. Petrology and geochemical characteristics

Komatiites exhibit compositional differences, stratigraphic and secular variations in major, minor and trace elements geochemistry depending on their location, their assemblage inside a location, and their age. The variety of data is impressive. We consider illustrative in the present general review to synthesize and summarize their geochemical features with respect to different types of rocks. But it is important to mention that, if describing in more details komatiites petrology and geochemistry, Archean komatiites should be better compared to younger high Mg rocks as present day basalts should not be compared to komatiites but to Archean basalts. Here we are reporting the main compositional information for four selected komatiitic locations corresponding to typical eras of the geological time-scale when komatiites are found, i.e at ~3.5 Ga (Archean Barberton Greenstone belt, BGB), ~2.7 Ga (Archean Pyke Hill, Abitibi greenstone belt, PH), 2.4 Ga (Paleoproterozoic Vetreny Belt, VB) and 89 Ma (Mesozoic Gorgona Island, GI). This compilation is regrouped in Tables 2, 3, and 4.

Komatiites often exhibit textural and compositional layering within individual flows which result from their unique fluid properties and composition (Huppert and Sparks, 1985). Many komatiites show unusual and spectacular textures, known as spinifex textures. Such textures are defined by large skeletal, plate-like olivine crystals in a finer-grained groundmass (generally of clinopyroxene, chromite, and glass). The crystals display parallel or randomly oriented grouping. The texture is explained by a magmatic quench crystallization effect promoted by rapid cooling of melt with low nucleation rate and high growth rate of crystals at a large degree of supercooling (Donaldson, 1982).

Komatiitic basalts are dominated by pyroxene, with less plagioclase. Olivine is only present where MgO is more than 12 wt. % (Arndt and Nisbet, 1982). Ultramafic komatiites are composed mainly of olivine, with interstitial pyroxene and little or no feldspar. Pyroxenes are augite, magnesian pigeonite and bronzite (Arndt *et al.*, 1977). On the Earth, magnesian pigeonite is common only in komatiite lavas and may be a defining characteristic.

Komatiite compositions are closer to that of the mantle than that of primitive basalt (Table 2). Komatiites are classified into two main types distinguished by their major and trace element contents. The aluminium-depleted komatiites are found only in the oldest records like Barberton (~3.5 Ga), and the aluminium-undepleted komatiites in the late Archean (e.g. Munro-type, Pike Hill, 2.7 Ga) and Mesozoic (e.g. Gorgona Island, 60 - 87 Ga). Nonetheless, some komatiites of the Barberton formation present Al-undepleted compositions. The Al-undepleted Barberton komatiites are characterized to be chondritic-like with $Al_2O_3/TiO_2$ of ~15 - 18 and $CaO/Al_2O_3$ ~1.1 - 1.5 (Chavagnac, 2004 -refer to Table 2), whereas the Al-depleted have low $Al_2O_3/TiO_2$ ratios (~ 8 - 11) and high $CaO/Al_2O_3$ ratios (~ 1.1 - 1.6). Munro-type presents also higher, near chondritic $Al_2O_3/TiO_2$. Barberton-type komatiites have moderately high levels of incompatible trace elements, whereas Al-undepleted komatiites (Munro-type and the Cretaceous komatiites from Gorgona Island) are typically depleted in incompatible trace elements. Rare Earth element (REE) patterns show depletion of the heavy REE (high Gd/Yb) for Barberton type komatiites, while Munro-type are depleted in lighter elements and have near-chrondritic ratios of the middle to heavy elements (Arndt, 2003, and see Table 3).

[TABLE 2]

[TABLE 3]

## 4.1. Komatiites and ore forming processes

Another interesting aspect of  komatiitic magmas is their importance in ore forming processes. They are among the few mafic/ultramafic lavas types that are S-undersaturated at the time of magma formation and do not reach S-saturation until a late stage in their ascent from the mantle. This S-undersaturation is due to the high temperature of these magmas produced by large degrees of partial melting of upper mantle source regions that were already depleted in S through earlier partial melting events. S-saturation of magmas leads to depletion in the chalcophile metals, in contrast komatiitic magmas retain the full complement of chalcophile elements (including Fe, Co,

Ni, Au, Cu Tl, Bi and PGEs) that the magma derived from the mantle source. When they become S-saturated they may form sulphides, strongly enriched in Ni, Cu, Au, PGEs and other chalcophile elements, which may accumulate to form ore deposits directly or be dispersed in the komatiites (Keays, 1995). Platinum-group element (PGE) abundances including Os, Ir, Ru, Pt, Pd have been reported in komatiites (Table 4). PGEs are present at higher levels in komatiites than in basalts making komatiites better probes of mantle PGE abundances (Puchtel and Humayun, 2000). In accordance with these authors, the spinifex-textured (MgO = 25 - 28%) and cumulate (MgO = 34 - 37%) komatiites are moderately enriched in Pt and Pd relative to Os and Ir with $(Pt/Os)N = 2.5\pm0.4$, and exhibit chondritic $(Os/Ir)N = 0.98\pm0.06$ ratios. Generally it is observed that, in ultramafic magmas, PGEs are very strongly and roughly equally enriched in any sulphide or metallic minerals that are present. They can provide unique information on the important role that sulfur plays during magmatic processes, a role still ignored by most petrochemists. It has been observed in many geological shields (e.g. Yilgarn, Perseverance, Pilbara, Zimbabwe, Kambalda, Abitibi terrains) that abundant komatiites host magmatic sulphide deposits rich in nickel (Barnes *et al*., 1995; Moore *et al*., 2000; Lahaye *et al*., 2001; Barnes, 2004; Hill *et al*., 2004;) and show high content of Ni. Almost all of the largest known concentrations of komatiite-hosted Ni sulfide deposits (but also Fe-Cu-PGE deposits) formed during the Archean. However, some komatiites of Permian-Triassic age (Northwestern Vietnam) are also associated with Ni-Cu-(PGE) bearing deposits (Glotov *et al*., 2001). Magmatic Ni-Cu sulphides strongly influence the precious metal contents of komatiites because they are greatly enriched in these metals relative to crustal rocks (Keays, 1982). Most komatiite-associated magmatic Ni-Cu-(PGE) sulfide deposits formed from sulfide undersaturated magmas and are interpreted to have formed in dynamic lava channels or magma conduits by incorporation of crustal sulfur. They commonly exhibit geochemical and isotopic evidence of crustal contamination (e.g., Th-U-LREE enrichment, negative Nb-Ta-Ti anomalies) and chalcophile element depletion on the scale of individual cooling units (Lesher *et al*., 2001). Thermal erosion and incorporation of sulphur-rich sea-floor sediments have been proposed as a mechanism by which the komatiites were brought to sulphide saturation (Huppert *et al*., 1984; Groves *et al*., 1986; Lesher and Groves, 1986).

Among the chemical analyses for determining PGE abundance in komatiites, osmium has been studied in detail, particularly its isotopic composition. As with modern plumes, the sources of Archean and Proterozoic komatiites exhibit a large range of initial $^{187}Os/^{188}Os$ ratios. Most komatiites are dominated by sources with chondritic Os isotopic compositions (e.g. Song La, Norseman-Wiluna, Pyke Hill, Alexo), though some (e.g. Gorgona) derive from heterogeneous sources (see Table 4). Some komatiites are enriched in [186]Os and [187]Os (Brandon *et al*., 2003; Gangopadhyay *et al*., 2003). The coupled enrichments of [186]Os/[188]Os and [187]Os/[188]Os are very similar to those displayed by the Hawaiian and Siberian plumes. Such enrichments could originate from the addition of ancient hydrothermally altered or metalliferous sediments into the source of plumes (Ravizza *et al*., 2001). But this is contradicted by mixing models and another explanation for the Os isotopic variations involves Os transfer from the outer core to the lower mantle in the late Archean (Puchtel *et al*., 2001; Brandon *et al*., 2003).

[TABLE 4]

Finally, one of the last attractive discoveries about komatiites has been the report of diamonds (Capdevila *et al*., 1999). Abundant diamonds ranging from microdiamonds size up to 4 mm were found in a volcaniclastic komatiite from the Proterozoic Dachine island arc in French Guiana, South America. This recent discovery was quite unexpected as the tectonic setting is distinct from that of all other currently exploited diamond deposits. It places significant constraints on the origin of komatiite magmas and the manner in which they interact with hydrated mantle in subduction zones. Capdevila *et al.* (1999) proposed that a primary, anhydrous komatiite magma formed by deep melting, then penetrated hydrated lithosphere beneath the ancient island arc where it collected both water and diamonds. As komatiite magma interacted with the relatively cool hydrated base of the mantle wedge, it become hydrous, its temperature and density decreased dramatically and it was ejected to the surface, bringing with it diamonds. The discovery implies also that some komatiites must have originated at depths of ~250 km or greater. It has been proposed that diamonds can be natural time capsules, preserving information about the cycling of sulfur between Earth's crust, atmosphere, and mantle some 3 billion years ago (Farquhar *et al.*, 2002). These authors report that diamonds from a region in Botswana, Africa contain a distinctive ratio of three isotopes of sulfur. The presence of this ratio indicates that the sulfur in these diamonds went through a nearly complete geochemical cycle. Thus, diamonds are valuable crystals through which geologists and atmospheric chemists can peer to gain insights into the Earth's atmosphere as it existed billions of years ago.

Whereas the petrological features of komatiites have been the purpose of a wide range of studies, relatively little is known about their volatile abundances because of significant alteration of rocks and lack of fresh glasses. Despite claims to the contrary, some komatiites display evidence of being volatile-bearing like hydromagmatic amphiboles (Stone *et al.* 1997), and degassing structures i.e. segregation structures and vesicles/amygdales (Beresford *et al.*, 2000). Studies of melt inclusions in olivines provide information about volatile contents of komatiites (McDonough and Ireland, 1993; Woodhead *et al.,* 2005). For example, melt inclusions found in nearly fresh Belingwe komatiites contain up to 1.1 wt% of water, detectable $CO_2$ (up to 200 ppm), sulfur (500-750 ppm), chlorine (400-700 ppm) and Cl with $Cl/K_2O$ ratios ~0.8 – 1.5 (Woodhead *et al.,* 2005).

In many komatiite sequences worldwide, there is a strong spatial correlation between volatile-rich phases or vesicles/amygdales, and different styles of komatiite-hosted Ni-Cu-(PGE) sulphide mineralisation. However, it is unclear whether the volatile content of mineralised sequences plays a role in the genesis of komatiite-hosted NiS mineralisation, either in the control of sulphur solubility or/and in the concentration of sulphide blebs (Fiorentini *et al.*, 2005).

# 5. Komatiites in the Solar System

As previously described, komatiites are defined as a primitive volcanic material of the early Earth, mainly present during the Archean era. But the possible existence of komatiites (or rocks with komatiite-like compositions) in other planetary bodies of our Solar System has also been proposed. We briefly illustrate it in the following part. However, it is important to stress that many ideas reviewed here, are still undoubtedly speculative, and should be presently considered as plausible hypotheses until further "real" in situ geochemical analyses can be performed.

## 5.1. The Moon

Williams *et al.* (2000b) have calculated lava liquid viscosity as a function of lava temperature and composition for several lavas. After comparison, they have found that lunar mare lavas have much lower viscosities (2.3 Pa.s at 1200°C) than modern, terrestrial tholeiitic basalts (53.8 Pa.s at 1200°C) and the closest terrestrial analogs to lunar lavas appears to be ancient Precambrian komatiitic lavas (with 4.4 Pa.s at 1200°C). To study lunar lava flows, the same authors have adapted the emplacement and thermal erosion model of Williams *et al.* (1998) used for analyzing the erosional potential of submarine komatiite lavas. Indeed komatiitic material presents lavas properties that may possibly occur in some lunar lavas such as low-viscosity (Hulme, 1973; Carr, 1974), turbulent emplacement (Hulme, 1973, 1982), thermal erosion (Carr, 1974; Williams *et al.*, 2000b) and low thermal conductivity (Williams *et al.*, 1999a). The new model of Williams *et al.* (2000b) indicates that lunar basalts could have erupted as turbulent flows on the Moon. Therefore lunar basalts, beside the resemblance in composition to terrestrial komatiite lavas, may have flowed in a similar way as komatiites. So if we fully understand the volcanology of komatiite, this might help in interpreting lunar features. In the same way, perhaps the lunar examples might help us eventually to understand komatiites.

In addition, coexisting pyroxenes, pigeonite and augite, have been found in ~12 categories of lunar mare basalts as well as in Apollo 12 pigeonite basalts. These basalts are also magnesium-rich and therefore it is argued that these coexisting pyroxenes may have formed by rapid metastable crystallization of a supercooled liquid, a similar mechanism to komatiite crystallization on Earth (Reyes and Christensen, 1994).

Finally, it has been observed moreover that the lunar crust seen in the Apollo 16 highland breccias contains a primitive component which has been found to be a komatiite compositionally similar in major and minor elements to terrestrial komatiites (Ringwood *et al*., 1987; Wentworth and McKay, 1988). It strongly implies a corresponding similarity between the source regions of komatiites in the lunar interior and the Earth's upper mantle.

## 5.2. Mars

SNC meteorites have been observed to present some similarities to komatiites.

SNC meteorites, as terrestrial meteoritic debris dated at 1.3 Ga, have been classified as Martian suspects (McSween, 1985). Analyses of their mineralogy and MgO contents have shown that Shergotty, Nakhla and Chassigny meteorites resemble a lower-Mg komatiitic basalt, a higher-Mg komatiitic basalt and a peridotitic komatiite, respectively (Reyes and Christensen, 1994). Shergotites have coexisting augite and pigeonite like many mafic komatiites, Nakhla and Chassigny have coarse cumulate textures common to the basal portions of terrestrial komatiites (Treiman, 1988). Certain shergotites may be extrusive lavas or near surface intrusions (McSween and Jarosewich, 1983). Those extrusive shergotites are ultramafic pyroxene-rich lavas with coexisting pigeonite and augite, like komatiitic basalts, making early Precambrian basaltic komatiites on Earth a possible candidate for an analog of Martian lavas (Reyes and Christensen, 1994). Other works have also stressed this possibility (Baird and Clark, 1981; Mustard and Sunshine, 1995).

It is important to note that the degree of ease with which fluid komatiite lavas may be generated on Earth and Mars is different. On Earth, melting 40% of the mantle to get komatiite lavas is a difficult exercise because very high temperatures are needed (1450 to 1700 °C) to achieve such high melting ratios. It was mainly during the first half of Earth History, when the mantle was arguably a few hundred degrees hotter than today's, that komatiite lava was produced. However, on Mars, komatiite could be much more widespread. If the Martian mantle is indeed much richer in iron, this type of magma would be produced at lower melting ratios than on Earth, i.e at lower mantle temperatures. Komatiites might then be the dominant lava type we see at the surface, rather than its close cousin basalt (Frankel, 1996).

Observations and data from the Phobos 2 Imaging Spectrometer (ISM) for Mars (Baird and Clark, 1981; Mustard *et al.*, 1993; Reyes and Christensen, 1994) appear to support this hypothesis. The chemical and mineralogical composition of Martian fines indicates derivation from mafic to ultramafic igneous rock, probably rich in pyroxene (Baird and Clark, 1981). Also, according to Mustard *et al.* (1993), the Syrtis Major volcanic materials analyzed by ISM are dominated by pyroxene, may contain both pigeonite and augite, and have little olivine, which is consistent with terrestrial komatiitic lavas (Reyes and Christensen, 1994). Nevertheless, as for the lunar mare basalts, caution is recommended because the emphasis on pyroxene as exclusive criterion of komatiite affinity is controversial.

Although still speculative, the possibility that komatiitic lavas could erode channels on Mars is not excluded (see Baird and Clark, 1984). Most of the objections to lava as the main erosive agent on Mars are based on the observed characteristics of basalt (eruption at ~1200 °C and high viscosity). However,

Huppert *et al.* (1984) demonstrate that terrestrial Precambrian komatiitic lavas, erupting as very hot, highly fluid, turbulent flows, may be capable of eroding deep channels, melting and assimilating rocks over which they flowed. Therefore if lavas on Mars had a komatiitic origin, a water like fluid would have been available and would explain the channel formation and the lateral extent of Martian flows, and the absence of apparent outflow sedimentary deposits in the basins into which the large channels discharge (Baird and Clark, 1984).

## 5.3. Venus

As Ghail (2001) states "Venus might be the key to understand what the early Earth was like during the late Archaean and early Proterozoic".

Volcanic terrains are abundant and diverse on Venus. Data from Pioneer-Venus spacecraft, from earth-based observatories, from Soviet Venera landers and orbiters have revealed evidence for hot spots, crustal spreading, plains volcanism and volcanic edifices, orogenic belts, convergence and crustal thickening (Hess and Head, 1990). Hess and Head (1990) have made extended review, considerations and examinations on the composition of Venus, source material for melting, melting and magmatic processes and conditions. They have concluded that on the basis of the little information presently known about Venus, a wide range of magma types is possible on Venus depending on which of the magmatic environments is dominant and the depth of melting. In the case of mantle plumes and hot spot environments, they have proposed that the melts entrained in the hot core would resemble basaltic komatiites whereas the cooler parts of the plume generate MgO-rich tholeiites.

Venus apparently does not have an extensive continental or felsic crust, it is almost as massive as Earth and was as hot or even hotter. Based on these considerations and in relationship to the early Earth history and the particular properties of komatiites, the same authors have proposed that komatiite volcanism was and perhaps still is extensive on Venus. And it could be possible that komatiites played a significant role in the early evolution of the Venusian crust.

Some of the long and voluminous lava flows observed on the surface of Venus may possibly represent such komatiitic deposits.

Because liquid water is unstable at present Venusian surface conditions, the discovery of canali channels on Venus, thousand kilometers long was not predicted. The canali-forming agents are considered to be low-viscosity lavas that remain fluid for such long distances and possible speculated lavas could include komatiites and high-Fe-Ti lunar type basalts (see Gregg and Greeley, 1993).

Over fifty major flow fields have been recognized such as Mylitta Fluctus which covers an area of approximately 300 000 km$^2$. It has been proposed that the morphology and extent of the lava flows are consistent either with a komatiitic, sulfuric or carbonatitic rich volcanism (McMillan, 2005).

However, additional sample analysis missions are undoubtedly needed to distinguish various hypotheses on the nature, origin, physical properties of Venusian magmatism and volcanism. For example, there are still no direct compositional constraints, which limits the modelling for eruption and lava flow emplacement on Venus.

The ESA mission, Venus Express, one of whose scientific objectives is to study the Venusian surface properties, will perhaps supply scientific data that could shed light on some ambiguities. Specifically, the VIRTIS instrument, a visible infrared thermal imaging spectrometer, will provide unique information on surface temperature, mineralogy, chemical weathering, recent volcanic activity and earthquakes occurrence (Marinangeli *et al.*, 2004).

## 5.4. Io

Among many discoveries about Io, one of the most astonishing reported from the Galileo spaceprobe data of active volcanoes on Jupiter's moon, is that some of Io's lavas possess eruption temperatures of 1430 - 1730 °C (Keszthelyi and McEwen, 1997; McEwen *et al.*, 1998) greater than any lavas erupted on Earth today and possibly since the start of Earth's geological history.

One of the interpretations of high-temperatures hotspots on Io is the occurrence of ultramafic materials similar to terrestrial Precambrian komatiites. Indeed, because of its orthopyroxene spinifex and high MgO content resulting in a high liquidus temperature of 1611 °C, the 3.3 Ga komatiite in the Commondale greenstone belt of South Africa has been proposed (Williams *et al.*, 1999b) as a useful terrestrial analog for the Ionian lavas. Eruptions on Io have been modeled applying komatiitic analogs and a terrestrial ultramafic flow model (Williams *et al.*, 2000a). The model revealed that the komatiites of the Commondale greenstone belt, South Africa, are consistent with available Galileo data on the temperatures and composition of potential Ionian ultramafic materials.

If the assumption of superheated ultramafic melts on Io is further corroborated and confirmed, then one might reevaluate the idea that a hot Archean mantle may have generated superheated melts (Williams and Lesher, 1998).

## 5.5. Chondritic bodies

As it has been noted earlier in this paper, komatiites present similarities in composition with chondritic material, which rises the assumption of a meteoritic source for the production of komatiites.

Although only using chondritic trace element or Os isotope ratios to link komatiites and chondritic meteorites may be considered debatable, Puchtel *et al.* (2004a) indicate that many Os isotopic compositions ($^{187}$Os/$^{188}$Os ratios) and Re-Os ratios ($^{187}$Re/$^{188}$Os) of komatiites are comparable to ratios found in ordinary chondrites. Also in some komatiites, high siderophile element (HSE) abundances occur in relative proportions similar to those found in average enstatite chondrites (Puchtel *et al.*, 2004a), which could be consistent with one model of the accretion arguing that the HSE abundances in the terrestrial mantle were inherited from chondritic material of a late veneer.

Komatiites exhibit also Rb/La and Sr/Sm ratios with near-chondritic values (Arndt, 2003) and the low Al/Si ratios found in some komatiites may as well indicate a chondritic composition for the Earth's early Archean mantle (Francis, 2003).

# 6. Komatiites and implications in Astrobiology

Although basalts are major extrusive rocks in the Archean (De Witt and Ashwal, 1997), komatiites may also represent another compositionally rocky component of the primitive environments where possibly early life on Earth could emerge and evolve, and it's plausible that they may have played a role in the evolution of the primitive atmosphere and maybe in some changes in the compositions of the hydrosphere.

Molecular evidence is increasingly strong that the first living organisms existed around (or were related to) a hydrothermal system. This "hydrothermal" hypothesis is supported by a wide variety of arguments. The most accepted is that all the deepest branches of life share a common hyperthermophile character (Kandler 1992, 1994) existing at temperatures of 80 - 110 °C. The hypothesis is also plausible in terms of the tectonic, paleontological, and degassing history of the Earth (Hoffman and Baross, 1985). As well, many of the most ancient proteins appear to have a hydrothermal origin and key elements that make possible the existence of living communities. S, Fe, Mn, Zn, and perhaps Ni and Mg may have entered into life processes from an early hydrothermal or volcanic substrate, or from fluids in a volcanic ambiance (Nisbet and Flower, 1996). From this reasoning, early life on the Archean Earth may have colonized a wide variety of hydrothermal environments which would have included hydrothermal systems on mid-ocean ridges, around komatiite plume volcanoes and in shallow water settings. On the continents, plume komatiitic volcanism (comparable to Hawaii but lower and much larger) would have created wide low shields.

Elements like sulphur, iron, manganese and magnesium, all characteristic of hydrothermal environments, play crucial roles in many functions of life, such as photosynthesis (Barber and Anderson, 1994). Komatiite eruptions as in other hydrothermal systems like submarine basalt possibly may have first supplied these elements to life. A wide variety of trace elements like Zn, Cu, Se and other rare elements also used in essential processes of life, not easily accessible to the early unskilled biosphere, would have been introduced to living organisms most likely by hydrothermal systems and possibly around komatiite plume volcanoes (Nisbet and Fowler, 1996).

A particular astrobiological interest in komatiites may be related to its ability to host large nickel sulphide deposits (Barnes, 2004; Hill *et al.*, 2004) accumulating along the bases of the flows and potentially exposed on active volcanoes by erosion or earthquakes. This has led Nisbet and Fowler (1996) to propose that besides hydrothermal vents around subduction zones and mid-ocean ridges, komatiitic shield volcanoes could provide another possible localized source of sulphur associated with iron. Such element association has crucial role for some life processes. Iron sulfides, as well as nickel sulfide have shown to promote a number of relevant electron transfer reactions and synthesis reactions such as the reduction of $CO_2$, CO and $N_2$ (Schoonen *et al.,*

2004).

As well, the same authors have suggested that, in some cases, hydrothermal penetration of nickel sulphide deposits in komatiite flows could have provided local Ni-rich fluids. Environmentally, Ni is a relatively rare element, not especially concentrated even in most hydrothermal fluids, and it is quite difficult to imagine how the primitive organisms may have obtained it. One possibility is that nickel had a meteoritic origin. As well, it might have derived, not only from the classical concentrations of this element in peridotites in fracture zones, but also possibly locally from large komatiitic plume volcanoes. Such volcanoes would form a huge shield structure and would produce enormous lava flows, flowing widely and then cooling with vigorous and shallow hydrothermal systems. Many key proteins are Ni-based, for example, urease, a key part of the nitrogen cycle, is built around Ni. Given that early cells were probably not highly skilled at extracting metals from the environment, it is possible that Ni-enzymes used in basic cellular housekeeping evolved in a setting where Ni sulphides were abundant and obtruded into early cells, perhaps serving as sites for the first Ni-metal proteins. Based on those considerations, Nisbet and Fowler (1996) have raised the assumption that such settings could have been around komatiite hydrothermal systems that penetrated Ni sulphide deposits in komatiite flows. Nickel sulphide layers would have provided substrates for bacteria. Although Ni is widely present in basalts, and thus available to life from the basaltic crust rich in olivine, it is possible that an exposed nickel sulphide horizon could over a few years or decades support selective evolution of Ni-containing proteins. In this perspective, Nisbet and Fowler have also proposed that urease could have had a komatiite-based history. This is based on the double assumption: a) urease may have evolved in the presence of abundant carbon dioxide as $CO_2$ is required for the assembly of the nickel metallocentre (Park and Hausinger 1995), and b) one possibility of encountering such conditions could have been around an exposed nickel-sulphide substrate in an eroded or fault broken lava flow on the flanks of an extensive komatiite plume volcano (Nisbet and Fowler, 1996).

Another possible interesting astrobiological point regarding komatiites has been exposed by experimental works on komatiite weathering (MacLeod *et al.*, 1994; Kempe and Kazmierczak, 2002). They have performed dissolution experiments with pulverized unweathered komatiites in water exposed to ambient $CO_2$ pressure during several weeks (Kempe and Kazmierczak, 2002) or komatiites in standard seawater at different temperatures (MacLeod *et al.*, 1994). The results have shown a decrease of $Ca^{2+}$ concentration and an increase in the alkalinity of the resulting solution. The resulting fluids were invariably alkaline. Kempe and Kazmierczak have proposed that the presence of komatiites during weathering by $H_2CO_3$, HCl and $H_2SO_4$ acids could have promoted alkaline solutions.

Among the hypothesis for the origin of life on Earth, one strongly argues that life arose in aqueous solution as in lagoons, lakes and oceans (Hoffman and Baross, 1985). In order for life to being able to emerge and develop in such solutions, particular chemical conditions would have been required, like non acidic composition (avoiding for instance the hydrolysis of proteins), non-oxic solution (avoiding a rapid oxidation of the organic matter) and low $Ca^{2+}$ concentrations (for correct biochemical functioning of proteins, avoiding protein denaturation and avoiding the reduction of the phosphate concentration for the production of ATP) (Kempe and Kazmierczak, 2002). Consequently, one argument for early life on Earth is that alkaline environments could be more promising sites for biogenesis.

When considering the scenario of the prebiotic chemistry, komatiites could have contributed, in addition to other sources, to the supply of inorganics and organics in the primitive Earth. The predominance of high-temperature mafic (e.g. picrites) and ultra-mafic (e.g. komatiites) lava on early Earth would have favor effusive eruptions with high $H_2$ and CO contents in volcanic gases owing to CO - $CO_2$ and $H_2O$ - $H_2$ equilibria in magmatic gases. In addition, the amounts of CO and $H_2$ (and $NH_3$) increase if magmas were more reduced (fO$_2$ down to C$^0$ -, Fe$^0$ –bearing buffers) than on the present Earth. These buffers provide significant thermodynamic drive to form hydrocarbons below ~400°C (Richard, 2005). The best conditions for organic synthesis on early Earth are achieved in submarine Hawaiian-type eruptions of high temperature and/or reduced magmas. The hydrocarbons might be formed by Fisher-Tropsch type synthesis catalyzed by magnetite and/or Fe$^0$ present in solid volcanic products (Anderson, 1984). Fisher-Tropsch type reactions may have produced hydrophobic compounds. Such hydrophobic material would have formed a hydrophobic layer on the surface of the sea, which would have provided an environment thermodynamically more suitable than water for the concentration and polymerization of organic molecules fundamental to life, particularly amino acids and pyridine bases.

To examine the possible role of komatiites in the production of H$_2$ and in the consequent production of hydrocarbons, Richard (2005) has studied thermodynamically the hydrothermal alteration of ultramafic rocks at slow-spreading mid-ocean ridges. The results of the calculations indicate that H$_2$ is produced in sufficient amounts to lead to detectable amounts of organics produced by a Fisher-Tropsch type reaction. Therefore, as the author concludes, in Archean environments, the alteration of komatiites could also have led to important production of H$_2$ and abiogenic organic carbon, part of which may have been preserved as residual carbonaceous material in the hydrothermal cherts overlying the komatiites.

In addition to organic carbon, Mather *et al.* (2004) pointed out that Archean komatiite lava could have released enough thermal energy for producing fixed nitrogen (NO, $NO_x$). On early Earth, before the biological nitrogen fixation, the sources of fixed nitrogen needed for the emergence and sustainability of life were abiotically driven by energetic processes. Such energy sources include thunderstorm lightning (Nna-Mvondo *et al.*, 2001; Navarro-González *et al.*, 2001), coronae discharges (Nna-Mvondo *et al.*, 2005), post-impact plumes (Kasting, 1990), and volcanic processes (Navarro-González *et al.*, 2000). Based on measurements of NOx on the crater rim of masaya volcano, Nicaragua, Mather and collaborators (2004) proposed that hot magmatic vents could offer the opportunity for atmospheric nitrogen to be rapidly heated and to undergo thermally catalyzed reactions producing nitrogen oxides. They have estimated the nitrogen-fixation potential of different magma types for the prebiotic and present atmosphere. The results show that during the Archean komatiite lava would release thermal energy at greater rate that the basaltic and andesitic magmas, therefore leading to higher production rates of fixed nitrogen ($\sim 2 \times 10^9$ to $2 \times 10^{11}$ mol $yr^{-1}$). These rates are comparable to estimated production rates in the early Earth by volcanic lightning, thunderstorm lightning, and post-impact plumes.

Finally, there is a significant and exciting relationship of early Earth komatiites with copper mineralization, regarding the origin of life and the "peptide world". In accordance with Planckensteiner *et al.* (2004) the availability of copper(ii) ions in a prebiotic scenario would enable the formation of peptides and proteins, which are the basic components of living organisms. As previously defined, Cu is an element which is retained in komatiitic magmas. The recently discovered Collurabbie Ni-Cu-PGE prospect in the northeast Yilgarn Craton (Western Australia) is an unusual, PGE-enriched nickel sulphide komatiite deposit (Jaireth *et al.*, 2005) and the Permian-Triassic komatiite basalt complex in the Song Da rift, northwestern Vietnam, occurs in the axial part of this structure and includes komatiites, komatiitic basalts, olivine basalts, and subvolcanic bodies of dunite and plagioclase-bearing wehrlite hosting NiCuPGE sulfide ores (Glotov *et al.*, 2001).

## 7- Summary

The unique character and origin of komatiites make them excellent indicators of the early composition and development of the Earth's mantle. Considered as "primitive" lavas, a significant constituent of the Hadean crust, and beside basalt another component of Archean greenstone belts, komatiites are important for understanding the evolution of the early Earth. But they are also very interesting for astrobiological exploration. Komatiites are not exclusively related to the early Earth, as they may resemble some geological features of other planetary bodies of the Solar System, particularly on Io, and could help us to understand some extraterrestrial volcanisms. Such a prospect opens the idea that komatiites might be more common in the Solar System and also that the definition of komatiites as Archean rocks could perhaps further be extended more commonly to planetary volcanic material. Komatiites have been observed to present chondritic-like compositional characteristics, as some meteorites have been reported to be similar to komatiites for specific features in their structure and geochemistry. It has been suggested that komatiite genesis might be influenced by impacts of extraterrestrial objects. But this impact hypothesis is based on theoretical models, and clearly needs to be further explored to better assign such an origin. This is of particular interest as meteorites are proposed to be the building blocks from which all planets are made and because they tell us a great deal about the primary mineral composition of planets. Undoubtedly, the connection between (large) impacts and komatiite genesis needs to be further explored.

We tend to forget the importance of rocks and minerals in chemistry,

bio/geochemistry and consequently for early life on Earth. It's probable that komatiites may have participated in the production of hydrocarbons and fixed nitrogen in the early Earth. Volcanic rocks such as komatiites could be a habitat for early microbial life, since at the beginning of Earth's history komatiite plume volcanoes could provide possible sites for the evolution of various biological processes. For instance, perhaps metals like nickel could have been incorporated into biochemistry around a komatiite volcano. Recently, new evidence has been found for ancient microbial activity within Archean pillow lavas of the Barberton Greenstone Belt (Furnes *et al.*, 2004). These pillow lavas are predominantly komatiitic and basaltic. Such a modern view makes komatiites much more attractive rocks that may offer new perspectives in the astrobiological Sciences. In this context, their investigation should extend well over the geological and geochemical field. Also the understanding of their significance should not be restricted to early Earth history but studied and interpreted as well for other planets and planetary bodies of the Solar System.

# Acknowledgements

We really thank referees Dallas Abbott and Nicholas Arndt whose helpful and constructive comments substantially improved the original manuscript. We also are particularly grateful to the referees for specific useful scientific remarks regarding the comparison of komatiites to Archean basalts (Dallas Abbott comment), and the interpretation of lunar lavas (Nicholas Arndt comment) which were incorporated verbatim into the article.

Finally, we acknowledge the grant support of the National Institute of Aerospace Technique "Esteban Terradas" (INTA) – Centro de Astrobiología.

**Table 1.** Ages of komatiites from different locations (listed in the descending order).

| Age (Ga)[a] | Geological terrain | Location | Reference |
|---|---|---|---|
| 3.6 - 3.2 | Barberton Greenstone Belt | South Africa | Byerly *et al.*, 1996; Lowe and Byerly, 1999 |
| 3.5 - 3.2 | Commondale Greenstone Belt | South Africa | López-Martínez *et al.*, 1992 |
| 3.47 | Pilbara Craton, North Pole Dome | Western Australia | Brown *et al.*, 2004 |
| 2.97 | Rio Maria greenstone terrane | Southeast Pará, Northern Brazil | De Souza *et al.*, 1997 |
| 2.9-2.7 | Abitibi Greenstone Belt (Munro, Alexo, Tisdale, Boston, Dundonald, Pyke Hill, Marbridge Townships) | Ontario, Canada | Puchtel *et al.*, 2004b; Gangopadhyay and Walker, 2001; Arndt, 1976 |
| 2.8 | Kostomuksha Greenstone Belt | Baltic Shield | Puchtel *et al.*, 2001a |
| 2.7 | Belingwe Greenstone Belt | Zimbabwe | Nisbet *et al.*, 1987 |
| 2.7 | Yilgarn Craton (Black Swan, Kambalda, Forrestiana, Perseverance complex in Norseman-Wiluna greenstone belt) | Western Australia | Barnes, 2004; Lahaye *et al.*, 2001; Perring *et al.*, 1996; Barnes *et al.*, 1995 |
| 2.4 | Vetreny belt | Southeastern Baltic Shield | Puchtel *et al.*, 2001b, 1996 |
| 2.1 | Dachine deposit, Inini geenstone belt | French Guiana | Capdevila *et al.*, 1999 |
| 2.06 | Central Lapland Greenstone Belt, (Finnish Lapland) | Finland | Hanski *et al.*, 2001; Gangopadhyay, 2002 |
| 0.270 | Song Da zone | Northwestern Vietnam | Hanski *et al.*, 2004 |
| 0.089 | Gorgona Island | Colombia | Echeverria, 1980; Arndt *et al.*, 1997 |

[a] General dating error roughly ranges from ± 0.01 to ± 0.8.

**Table 2.** Major element abundances (wt.%) in komatiites located in different geological settings, in comparison with the earth's undepleted mantle (UM) and primitive basalt abundances (PB).

| Location[b] | BGB | PH | VB | GI | UM | PB |
|---|---|---|---|---|---|---|
| $SiO_2$ | 47.2 | $44.4 - 45.8$ | $45.4 - 53.6$ | $44.6 - 47.6$ | 45.1 | 44.2 |
| MgO | 31.4 | $26.8 - 34.7$ | $6.8 - 26.1$ | $21.9 - 14.2$ | 38.1 | 13.1 |
| $Fe_2O_3$ | — | $10.1 - 13.0$ | $10.8 - 13.4$ | $12.1 - 12.9$ | — | — |
| LOI | — | $6.1 - 9.1$ | $0.2 - 5.3$ | $2.8 - 5.2$ | — | — |
| $Al_2O_3$ | 3.4 | $5.1 - 7.2$ | $7.2 - 13.8$ | $13.1 - 10.6$ | 3.3 | 12.1 |
| CaO | 5.6 | $3.7 - 7.5$ | $7 - 11.6$ | $9.2 - 10.8$ | 3.1 | 10.1 |
| FeO | $11.2^{b}$ | — | — | — | 8.0 | 10.9 |
| $Na_2O$ | 0.07 | $0.24 - 0.50$ | $0.4 - 2.5$ | $0.68 - 1.4$ | 0.4 | 3.6 |
| $TiO_2$ | 0.3 | $0.25 - 0.35$ | $0.4 - 0.8$ | $0.4 - 0.7$ | 0.2 | 3.7 |
| MnO | 0.2 | $0.11 - 0.19$ | $0.16 - 0.19$ | $0.17 - 0.22$ | 0.1 | 0.2 |
| $K_2O$ | 0.04 | $0.02 - 0.09$ | $0.1 - 0.7$ | $0.02 - 0.05$ | 0.03 | 1.3 |
| $P_2O_5$ | 0.02 | $0.01 - 0.02$ | $0.06 - 0.2$ | $0.03 - 0.06$ | — | 0.8 |
| $H_2O$ | — | — | $0.4 - 0.8$ | 1.89 | — | — |
| Ref.[c] | (1) | (2) | (3) | (4) | (5) | (6) |

[b] Location: BGB (Barberton greenstone belt), PH (Pyke Hill, Abitibi Greenstone belt), VB (Vetreny Belt), GI (Gorgona Islang).

[c] Data sources in Tables 2, 3 and 4: (1) Parman *et al*., 2003; (2) Puchtel *et al*., 2004a; (3) Puchtel *et al*., 1997; (4) Kerr, 2005; (5) Francis and Oppenheimer, 2004; (6) Basaltic Volcanism Study Project, 1981; (7) Puchtel and Humayun, 2000; (8) Puchtel *et al*. (2001b); (9) Brandon *et al*., 2003.

**Table 3.** Minor and trace element abundances (ppm) in komatiites located in different geological settings[d].

| Location[b] | BGB | PH | VB | GI |
|---|---|---|---|---|
| Cr | 2920 | 2349 - 2887 | 283 - 3507 | 796 - 5614 |
| Ni | 1611 | 1330 - 2600 | 57 - 1138 | 464 - 956 |
| Zr | 19.1 | 11.9 – 17.6 | 34 - 71 | 18 - 33 |
| Y | 6.8 | 8.8 – 6.1 | 9 - 17 | 12 – 17 |
| Ga | — | 4.8 – 7.4 | — | 10 - 15 |
| Nd | 2.5 | 1.1 – 1.6 | 5 - 12 | 1.4 – 2.8 |
| Ce | 3.2 | 1.1 – 1.5 | 9 – 24 | 0.9 – 2.1 |
| Dy | 1.08 | 0.9 – 1.4 | 1.5 – 3 | 2.3 –2.6 |
| Gd | 0.9 | 0.7 – 1.1 | 1 – 3 | 1.5 –1.9 |
| Er | 0.6 | 0.6 – 0.9 | 0.9 – 2 | 1.4 –1.7 |
| Yb | 0.6 | 0.6 – 0.9 | 0.9 – 2 | 1.2 –1.5 |
| Sm | 0.8 | 0.5 – 0.7 | 1 – 3 | 0.7 –1.3 |
| Nb | 1.05 | 0.3 – 0.4 | 1 – 3 | 0.3 – 0.8 |
| La | 1.2 | 0.3 – 0.5 | 4 – 11 | 0.3 – 0.6 |
| Eu | 0.3 | 0.2 – 0.3 | 0.4 – 0.9 | 0.3 – 0.6 |
| Th | 0.1 | 0.03 – 0.04 | 0.7 – 3 | 0.02 – 0.03 |
| Ref.[c] | (1) | (2) | (3) | (4) |

[d] Notes b and c are the same as in Table 2.

**Table 4.** PGE abundances (ppb) in komatiites located in different geological settings[e].

| Location[b] | BGB | PH | VB | GI |
|---|---|---|---|---|
| $^{187}Os/^{188}Os$ | — | 0.1090±0019 | 0.1110±0.0013 | 0.1138±0.0032 |
| Os | 0.792 | 2.38 | 0.04 - 0.8 | 0.8 4 – 2.85 |
| Ir | 1.23 | 2.13 | 0.06 - 1.7 | 2.05 |
| Ru | 4.30 | 5.31 | 0.9 - 5.5 | 3.3 |
| Pt | 5.83 | 11.1 | 7.8 - 11 | 11.1 – 16.0 |
| Pd | 5.51 | 10.6 | 7.3 - 13.8 | 12 |
| Ref.[c] | (7) | (2) | (7), (8) | (4), (9) |

[e] Notes b and c are the same as in Table 2.